\documentclass[aps, amsmath,amssymb,prb,twocolumn,superscriptaddress]{revtex4-1}
\usepackage{graphicx,hyperref}
\usepackage[utf8]{inputenc}
\usepackage{amsmath,bm}
\usepackage{bbold}
\usepackage{color,soul}
\usepackage{xstring}
\usepackage{wasysym}
\usepackage{xspace}
\usepackage{amssymb}
\usepackage{time}
\usepackage{bbold}
\usepackage{subfigure}
\usepackage{multirow}
\usepackage{xspace}
\usepackage{url}

\def\<{\langle}
\def\>{\rangle}

\newcommand{\ve}[1]{\boldsymbol{#1}}

\usepackage{ulem}

\begin{document}

\title{ Gross-Neveu Heisenberg criticality: dynamical generation of  quantum spin Hall  masses  }

\author{\firstname{Yuhai} \surname{Liu}}
\affiliation{\mbox{Beijing Computational Science Research Center, Beijing 100193, China}}
\author{\firstname{Zhenjiu} \surname{Wang}}
\affiliation{\mbox{Institut f\"ur Theoretische Physik und Astrophysik, Universit\"at W\"urzburg, Am Hubland, 97074 W\"urzburg, Germany}}
\author{\firstname{Toshihiro} \surname{Sato}}
\affiliation{\mbox{Institut f\"ur Theoretische Physik und Astrophysik, Universit\"at W\"urzburg, Am Hubland, 97074 W\"urzburg, Germany}}
\author{\firstname{Wenan} \surname{Guo}}
\email{waguo@bnu.edu.cn}
\affiliation{\mbox{Department of Physics, Beijing Normal University, Beijing 100875, China }}
\affiliation{\mbox{Beijing Computational Science Research Center, Beijing 100193, China}}
\author{\firstname{Fakher F.} \surname{Assaad}}
\email{fakher.assaad@physik.uni-wuerzburg.de}
\affiliation{\mbox{Institut f\"ur Theoretische Physik und Astrophysik, Universit\"at W\"urzburg, Am Hubland, 97074 W\"urzburg, Germany}}
\affiliation{\mbox{W\"urzburg-Dresden Cluster of Excellence ct.qmat, Am Hubland, 97074 W\"urzburg, Germany}}
\begin{abstract}

We consider  fermions on a  honeycomb lattice supplemented by a spin invariant  interaction that  dynamically generates a quantum spin Hall insulator.  This  lattice model provides an instance of Gross-Neveu Heisenberg criticality, as  realized for example by the Hubbard model on the honeycomb lattice.   Using auxiliary field   quantum Monte Carlo  simulations we show  that  we can compute with unprecedented precision   susceptibilities of the order parameter.   In   O(N) Gross-Neveu transitions, the anomalous dimension of the bosonic mode grows as a function of N  such that  in the \textit{large}-N limit it is of  particular importance to consider    susceptibilities  rather than equal time correlations so as to minimize  contributions from the background. For the N=3   case, we  obtain $1/\nu=1.11(4)$, $\eta_{\phi}=0.80(9)$, and $\eta_{\psi}=0.29(2)$  for  respectively the correlation length exponent,  bosonic and fermionic anomalous dimensions.

\end{abstract}
\maketitle
\section{Introduction}
Fermionic quantum criticality is a long standing problem in the  domain of strongly correlated electron system  \cite{Hertz76,Millis93}.       In d-wave superconductors \cite{xu2020competing,Otsuka20} or in free standing graphene  \cite{Assaad13,Otsuka16,Toldin14}   the problem greatly simplifies.  Here,  the Fermi surface consists of a discrete set of points with   a linear  dispersion  relation at low energies  that can be captured by a Dirac equation  \cite{Neto_rev}.     Fermion criticality in these  systems refers to a set of phenomena such as  the  opening of a single  particle gap  \cite{Herbut09a} (mass generation),  or  nematic transitions   where  Dirac point meanders  \cite{Vojta00,Kim08}.

In mass generating  transitions  in two spatial dimensions,  one expects   emergent Lorentz symmetry \cite{Herbut09}. The field theory   corresponds to  Dirac fermions   supplemented by a  Yukawa term consisting   of a Dirac  mass   \cite{Ryu09} coupled to a   bosonic mode   described by a  $\phi^4$  theory \cite{Herbut09a}.    At the  Wilson-Fisher  fix-point,  the Yukawa  coupling is relevant and drives the system to a new so called Gross-Neveu  (GN) critical point.  In comparison to  Wilson-Fisher fixed points where the bosonic anomalous dimension is  small  \cite{Hasenbusch10,Vicari02,Kos16},   fermion quantum criticality in Dirac systems is characterized by a much larger one.  This can be understood intuitively  since  coupling to fermions provides new decay  channels for bosonic modes.   As noted in \cite{Toldin14}, this characteristic of the   GN  critical  points  potentially posses a numerical challenge.   If in  two spatial dimensions,   the  anomalous dimension of the bosonic mode is greater than unity  then,  the equal time correlations of this mode  will be dominated by the background.  On the other hand, critical fluctuations will become apparent in the  susceptibility.   In principle this should not  cause a problem since within  auxiliary field quantum Monte Carlo (AFQMC)  methods \cite{Blankenbecler81,Hirsch85,White89,Assaad08_rev} one can compute time displaced correlation functions and  hence susceptibilities.    To the best of our knowledge, it  turns out that  computing susceptibilities  for  Hubbard type models in the vicinity of  the critical point   is very noisy,  and  is  plagued by  rare configurations with anomalous fluctuations.     This  inhibits a  precise determination of this quantity and to  date  analysis of GN criticality in lattice systems \cite{Assaad13,Otsuka16,Toldin14,Lang18,Huffman19,He17,LiuY20} are based on equal time correlations of the critical bosonic mode.

In Ref.~\cite{Liu18},  we have introduced a model   with an SU(2) spin symmetry that shows   a  transition from a Dirac semi-metal (DSM) to a quantum spin Hall  (QSH) insulator.   As  conjectured in Ref.~\cite{Herbut09a}   this transition is expected to belong to the same universality class  as  that of the Hubbard model on the Honeycomb lattice. Remarkably,  our AFQMC  implementation presented in Ref.~\cite{Liu18}  does not  suffer from the  aforementioned anomalous fluctuations  of the critical bosonic modes.   We are hence in a position to compute the  susceptibility  and extract critical  exponents using this quantity.     The main result of  paper reads:
\begin{equation}
\label{Main_res.eq}
1/\nu=1.11(4), \eta_{\phi}=0.80(9), \text{ and } \eta_{\psi}=0.29(2)
\end{equation}
for the  exponents of the  (2+1)-dimensional GN-Heisenberg universality class at $N_f=2$   four component fermion fields
 akin to graphene.
Here   $\nu$  is the correlation length exponent   and $\eta_{\phi}$ ($ \eta_{\psi})$  the bosonic (fermionic) anomalous dimension.

The  article  is organized as follows.  In the next section,  we define the model and the AFQMC approach. In Sec.~\ref{sec:results},  we  discuss our QMC results using a  crossing-point analysis based on the time displaced correlations.    Adopting  this analysis scheme,   corrections to scaling are taken into account.
 In Sec.~\ref{sec:conclusions} we compare our results to previous estimates  and provide concluding remarks.    We  have included two appendices. In Appendix ~\ref{AppendixA}  we    compare the quality of our susceptibility data to those of the generic Hubbard model on the honeycomb lattice.    In Appendix~\ref{AppendixB}    we provide a detailed symmetry based understanding of the single particle Green function, that is used to compute the fermion anomalous dimension.

\section{Model and method}

\begin{figure}[b]
  \includegraphics[width=0.45\textwidth]{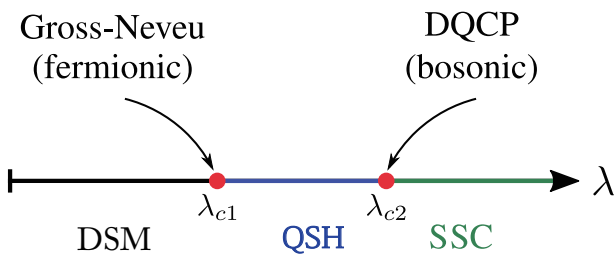}
  \caption{\label{fig:Phase_diagram}
     Schematic ground-state phase
    diagram with DSM, QSH, and SSC phases.  The DSM-QSH  transition belongs to the Gross-Neveu Heisenberg universality class.  The QSH-SSC is an example of a monopole free deconfined quantum critical point (DQCP) \cite{Liu18}.
 }
\end{figure}

We consider a model of Dirac fermions in $2+1$
dimensions on the honeycomb lattice with Hamiltonian
\begin{eqnarray}\label{Eq:Ham}
 \hat{H}  & = &- t  \sum_{ \langle \bm{i}, \bm {j} \rangle }  \hat{\ve{c}}^{\dagger}_{\bm{i} } \hat{\ve{c}}^{\phantom\dagger}_{\bm{j}}
   -\lambda \sum_{\varhexagon}  \left( \sum_{\langle \langle \bm{i} \bm{j} \rangle \rangle  \in \varhexagon }   \hat{\ve{J}}_{\bm{i},\bm{j}} \right)^2 \, ,
   \nonumber \\
   & &  \hat{\ve{J}}_{\bm{i},\bm{j}}  =  i \nu_{ \bm{i} \bm{j} }
 \hat{\ve{c}}^{\dagger}_{\bm{i}} \bm{\sigma}
\hat{\ve{c}}^{\phantom\dagger}_{\bm{j}}+\bm{H.c}  \, .
\end{eqnarray}
The spinor
$\hat{\boldsymbol{c}}^{\dag}_{\ve{i}} =
\big(\hat{c}^{\dag}_{\ve{i},\uparrow},\hat{c}^{\dag}_{\ve{i},\downarrow}
\big)$ where $\hat{c}^{\dag}_{\ve{i},\sigma} $ creates an electron at lattice
site $\ve{i}$ with $z$-component of spin $\sigma$. The first term
accounts for nearest-neighbor hopping. The second term is a  hexagon
interaction involving next-nearest-neighbor pairs of sites and phase factors
$\nu_{\boldsymbol{ij}}= - \nu_{\boldsymbol{ji}} = \pm1$ identical to those of  the Kane-Mele model \cite{KaneMele05}.  That is, assume that the honeycomb
lattice spans the x-y plane,  and let  $\ve{r}$  be the nearest neighbor site
common to  next nearest neighbor   sites $\ve{i}$ and  $\ve{j} $, then
\begin{equation}
	   \nu_{\ve{ij}} =       \text{ sgn } \left[  \left( \ve{i}  - \ve{r} \right)  \times  \left( \ve{r}  - \ve{j} \right)  \right]    \cdot  \ve{e}_z.
\end{equation}
    Finally,
$\boldsymbol{\sigma}=(\sigma^x,\sigma^y,\sigma^z)$ corresponds to  the vector  of Pauli
spin matrices.
Since   $ \hat{\ve{J}}_{\bm{i},\bm{j}} $    transforms as a vector  under SU(2)  spin rotations, the model possesses global SU(2) spin symmetry.

The ground state phase diagram as a function of $\lambda$ presented in Ref.~\cite{Liu18} is briefly summarized in Fig.~\ref{fig:Phase_diagram}.  As a  function of $ \lambda/t $ (we set $t=1$),we observe three phases:  a Dirac semi-metal
(DSM) for $ \lambda < \lambda_{c1}$; a quantum spin Hall (QSH) insulator   for $ \lambda_{c1} < \lambda < \lambda_{c2} $;   and an s-wave superconductor (SSC) at $ \lambda > \lambda_{c2} $. The DSM and QSH states are separated by a Gross-Neveu Heisenberg phase transition at $\lambda_{c1} \approx 0.0187$; the QSH and SSC states are separated by a deconfined quantum critical point (DQCP)  \cite{Senthil04_1,Senthil04_2,Grover08} at $ \lambda_{c2} \approx 0.0332$.  Here  and in  comparison to Ref.~\cite{Liu18}   we  focus on the critical behavior of  Gross-Neveu Heisenberg  transition.  We will  provide results on larger lattice  sizes  (up to  $L=24$) and   determine
the correlation length exponent, bosonic and fermionic anomalous dimensions.

\begin{figure}[b]
  \includegraphics[width=0.45\textwidth]{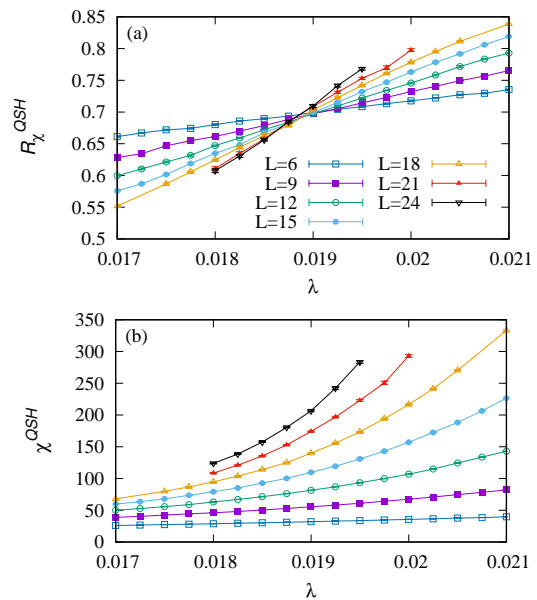}
  \caption{\label{fig:chiandratio}
   (a) Correlation ratio and  (b) susceptibility of the spin-orbit coupling order parameter for different system sizes across the DSM-QSH phase transition. }
\end{figure}
\begin{figure}[b]
  \includegraphics[width=0.45\textwidth]{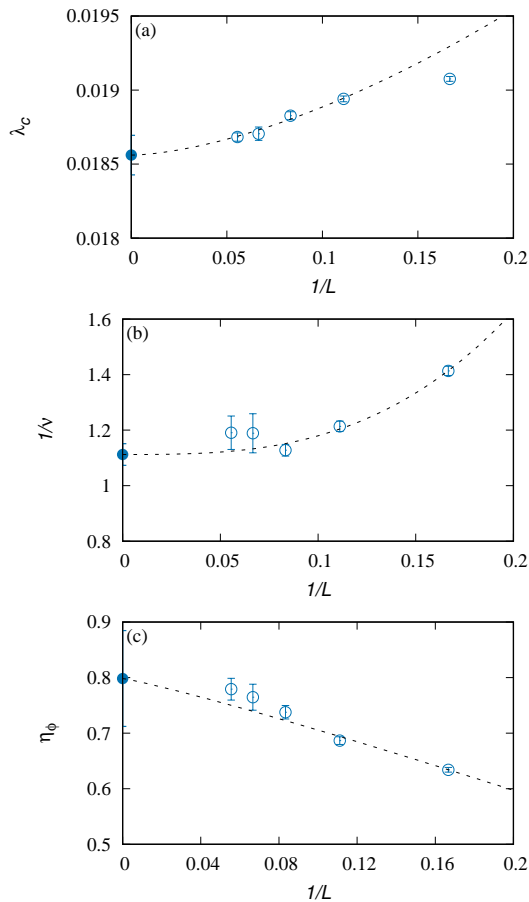}
  \caption{\label{fig:critical}
  (a)  $\lambda_c$ as a function of system size is obtained from a crossing point analysis  of the correlation ratio of Eq.~\ref{eq:correlation_ratio} for $L$ and $L+6$.  (b)   Correlation length exponent as a function of system size  as obtained from Eq.~\ref{eq:nufromcorrelationratio}.   (c)   Bosonic  anomalous dimension as obtained from Eq.~\ref{eq:eta_phi}.
From the fits (see  text) we  obtain $\lambda_{c}=0.0186(2)$, $1/\nu=1.11(4)$, and $\eta_{\phi}=0.80(9)$  in the large system size limit.
}
\end{figure}

The model described by Hamiltonian (\ref{Eq:Ham}) is investigated   with the  Algorithms for Lattice Fermions  (ALF) \cite{ALF_v1,alfcollaboration2021alf}   implementation  of finite temperature auxiliary-field quantum Monte Carlo (AFQMC)  \cite{Blankenbecler81,Hirsch85,White89,Assaad08_rev}.    Since the interaction is written in terms of squares of single body operators, the model is readily   implemented in the ALF-library. We   consider values of  $\lambda>0$ such that for a  given instance of Hubbard-Stratonovitch fields,  time reversal symmetry is present.  This has  for consequence that   the eigenvalues of the fermion matrix occur in complex conjugate pairs  \cite{Wu04}.  Hence no sign problem occurs.  Note that since   adding a chemical potential does not break   time reversal symmetry,   finite dopings can also be considered \cite{wang2020dopinginduced}.
For the details of the implementation of the  algorithm, we refer the  reader to   Ref.~\cite{Liu18}.
In the following, we used $t = 1$ as the energy unit and simulated half-filled lattices with $L \times L$ unit cell with periodic boundary conditions.
For the  numerical simulations presented here, we have used a symmetric  Trotter  decomposition  (see Ref.~\cite{alfcollaboration2021alf})     so as to ensure hermiticity  of the  imaginary time propagation.   For the  imaginary time step  we have chosen,  $\Delta_{\tau}= 0.2$ and as appropriate  for Lorentz invariant systems have     carried out an inverse temperature $\beta = L $ scaling  analysis.

One key technical point of this study is that our specific implementation  allows for the calculation of  the order parameter susceptibility with unprecedented precision.  In comparison to the Hubbard model on the honeycomb lattice, we  show in Appendix~\ref{AppendixA} that we  do not suffer from rare configurations with anomalous fluctuations   when computing this quantity.

\section{Results}
\label{sec:results}
\subsection{Order parameter}
\begin{figure}[b]
  \includegraphics[width=0.45\textwidth]{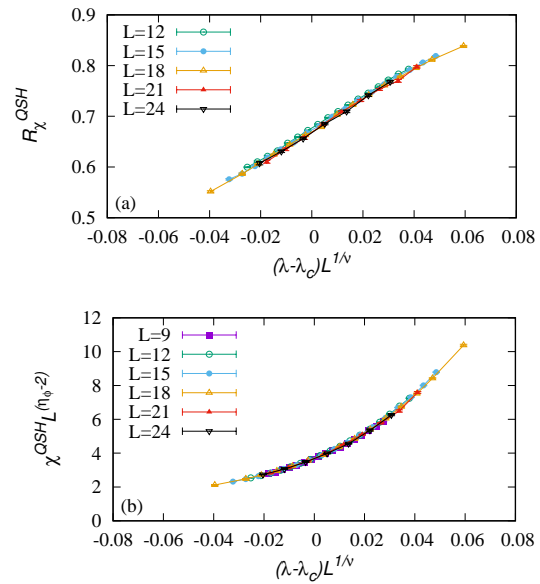}
  \caption{\label{fig:datacollapse}
 As a cross-check for   our determination of the critical point and exponents  we provide  a data collapse with $\lambda_{c}=0.0186 $, $1/\nu=1.11$, and  $\eta_{\phi}=0.8$   for (a) the correlation ratio and (b) the QSH susceptibility.  }
\end{figure}
\begin{figure}[b]
  \includegraphics[width=0.45\textwidth]{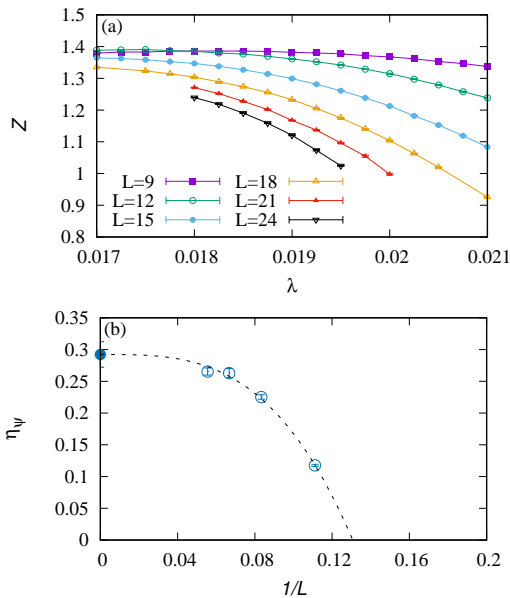}
  \caption{\label{fig:SPGreen}
   (a) Monte Carlo estimate of  $Z$ as defined in Eq.~\ref{Eq:Zr} (a)   Size scaling of the fermionic anomalous
   dimension as obtained from Eq.~\ref{eq:eta_psi}.  In the large system  size limit, we obtain:   $\eta_{\psi}=0.29(2)$.  }
\end{figure}
The DSM-QSH transition involves the breaking of an SU(2) spin rotation symmetry and is expected to be in the Gross-Neveu Heisenberg universality class for $N_f = 2$ four component Dirac fermions (two sublattices, two Dirac cones,  and spin $\sigma=\uparrow,\downarrow$ ).
 The local vector order parameter takes the form of the spin-orbit coupling,
\begin{equation}
\label{eq:O_QSH}
\hat{\boldsymbol{O}}^{QSH}_{  {\ve r},\langle\langle\ve{\delta},\ve{\delta}'\rangle\rangle}  =
\mathrm{i}\hat{{\ve c}}^{\dagger}_{  {\ve r}, \ve{\delta}} \ve{\sigma}
\hat{{\ve c}}^{}_{ {\ve r}, \ve{\delta}'}   +  \text{H.c.},
\end{equation}
where $\ve{r}$   labels  a unit cell or equivalently  a hexagon, $\langle\langle {\ve{\delta}}, {\ve{\delta}'}\rangle\rangle$ corresponds to  next-nearest neighbour pairs with legs $\bm{\delta} $ and $\bm{\delta}' $ of the corresponding hexagon. Because this order parameter is a lattice regularisation of the three QSH mass terms in the Dirac equation  \cite{Ryu09}, long-range order implies a mass gap. To study this phase transition, we use susceptibilities rather than equal-time correlations to suppresses background contributions to the critical fluctuations.
The associated time-displaced correlation functions of the spin-orbit coupling order parameter read
\begin{eqnarray}
&&S^{QSH}_{\langle\langle\ve{\delta},\ve{\delta}'\rangle\rangle \langle\langle\ve{\delta}'',\ve{\delta}'''\rangle\rangle}({\ve k}, \tau)=\nonumber\\
&&\frac{1}{L^2}\sum_{ {{\ve r},{\ve r'}} } \langle \hat{\ve{O}}^{QSH}_{ {\ve r},\langle\langle\ve{\delta},\ve{\delta}'\rangle\rangle}(\tau)\cdot \hat{\ve{O}}^{QSH}_ { {\ve r'},\langle\langle\ve{\delta}'',\ve{\delta}'''\rangle\rangle}(0) \rangle
{\ve e}^{i {\ve k} ({{\ve r}- {\ve r'}}) }  \;.
\end{eqnarray}

Here $\tau$ is the imaginary time.  Since our model enjoys  an  SU(2) spin rotation symmetry and
$ \hat{\boldsymbol{O}}^{QSH}_{  {\ve r},\langle\langle\ve{\delta},\ve{\delta}'\rangle\rangle} $ transforms as a vector under global rotations, we can neglect the background terms.
We  define the susceptibility as
\begin{equation}
   \chi^{QSH}({\ve k}) = \Lambda_1 \left( \mbox{$\int_{0}^{\beta}$}  d\tau S^{QSH}_{\langle\langle\ve{\delta},\ve{\delta}\rangle\rangle, \langle\langle\ve{\delta}'',\ve{\delta}'''\rangle\rangle}({\ve k},\tau)  \right) \,,
\end{equation}
where, $\Lambda_1()$ indicates the largest eigenvalue of the corresponding $6 \times 6$ matrix spanned by the $\langle\langle\ve{\delta},\ve{\delta}'\rangle\rangle$ and
$\langle\langle\ve{\delta}'',\ve{\delta}'''\rangle\rangle $  indices corresponding to the six   next nearest neighbor bonds of a hexagon.  The corresponding renormalisation-group invariant correlation ratio   \cite{Kaul15}  reads:
\begin{equation}
\label{eq:correlation_ratio}
R^{QSH}_{\chi} =1-\frac{\chi^{QSH}({\ve k}=\Delta {\ve k})}{\chi^{QSH} ({\ve k}=0)}.
\end{equation}
The ordering wave vector   corresponds to  ${\ve k}=0$, and on an $L \times L $ lattice   with periodic boundary conditions,  $|\Delta {\ve k}|=\frac{4\pi}{\sqrt{3} L }$.   In the thermodynamic limit  $R^{QSH}_{\chi}\rightarrow1$  ($R^{QSH}_{\chi}\rightarrow 0 $ )  in the ordered (disordered) phase and  corresponds  to a  renormalization group invariant quantity   %
\begin{equation}\label{eq:correlationratio}
R^{QSH}_\chi=f_{R}\left( L^z/\beta,\left( \lambda - \lambda_c \right)L^{1/\nu}, L^{-w}  \right),
\end{equation}
at the critical point.

Here $\beta$ is the inverse  temperature, $z$ the dynamical critical exponent, $\nu$ the correlation length exponent and $\omega$ the leading correction-to-scaling exponent.  We will assume conformal invariance and set  $z = 1$  and   $\beta = L$.  Hence up to corrections to scaling, $R^{QSH}_{\chi}  $   should show a crossing
point at $\lambda = \lambda_c$.  This is clearly seen  in   Fig.~\ref{fig:chiandratio}(a).  In Fig.~\ref{fig:chiandratio}(b)  we present  the bare data,  that   support a divergence of  the susceptibility beyond the crossing point  of the correlation ratio. In particular,  in the ordered phase,  we expect the correlation length to  diverge exponentially  with  inverse temperature \cite{Chakravarty88}.   For our  $\beta=L$ scaling   it  will hence exceed the size of the system and we expect the susceptibility to  scale as the   Euclidean  volume $ \beta L^2$  in the large volume limit.

We locate the critical point with the crossing point method.  Aside from a polynomial interpolation of the data as a function of $\lambda$ for each $L$, this analysis does not require any further fitting, and by definition, converges to the correct critical coupling with leading finite-size corrections given by   $L^{-\omega - \frac{1}{\nu}}$.   Figure~\ref{fig:critical}(a)  plots the finite-size estimate, $\lambda_c(L)$, corresponding  to the crossing point of $R_{\chi}^{QSH}$ for $L$ and $L+6$. Extrapolation to the thermodynamic limit yields $\lambda_c=0.186(2)$ and $\lambda_c(L)=\lambda_c+a_1L^{-\omega_1}$ with $\omega_1=1.6(5)$. Here $\omega_1$, also include $\omega_2$, $\omega_3$ and $ \omega_4$ corresponding to the correlation length exponent, bosonic and fermionic anomalous dimensions respectively in the later part should be considered  as 'effective' exponents that change with the range of system sizes considered, which  becomes the leading correction exponent only for very large sizes.

We  compute the  correlation length exponent, $\nu$, at crossing points of the  correlation ratio via
\begin{equation}\label{eq:nufromcorrelationratio}
  \frac{1}{\nu^{QSH}  (L) } =
  \left. \frac{1}{\log{r} }
    \log \left( \frac{ \frac{d} {d\lambda} R^{QSH}_\chi\left( \lambda, r L \right)  }{\frac{d} {d\lambda} R^{QSH}_\chi\left( \lambda,  L \right)  } \right)    \right|_{\lambda = \lambda_c(L)}
  \end{equation}
Here $r=\frac{L}{L+6}$.   The data of Fig.~\ref{fig:critical}(b)    supports $1/\nu=1.11(4)$    and $1/\nu(L)=1/\nu+b_1L^{-\omega_2}$ with $\omega_2=2.9(8)$.

To estimate the bosonic anomalous dimension   we consider  the susceptibility,
\begin{equation}\label{eq:susceptibility}
\chi^{QSH}(\ve{k} = 0 )= L^{2-\eta_{\phi}}f_{\chi}\left( L^z/\beta,\left( \lambda - \lambda_c \right)L^{1/\nu}, L^{-w}  \right),
\end{equation}
at   criticality such that
\begin{equation}
\label{eq:eta_phi}
\eta_{\phi}(L,rL)=2-\frac{1}{\ln(r)}\ln \left( \frac{\chi^{QSH}(\lambda_c(L),rL)}{\chi^{QSH}(\lambda_c(L),L)}\right).
\end{equation}
Again  $r=\frac{L}{L+6}$,   and $\lambda_c(L)$ refers to the  size resolved crossing point of the correlation ratio.
The data of Fig.~\ref{fig:critical}(c)    supports
 $\eta_{\phi}=0.80(9)$ with $\eta_{\phi}(L)=\eta_{\phi}+c_1L^{-\omega_3}$  and  $\omega_3=1.4(6)$.

Finally, we check the critical point and exponents  by collapsing the  data on the basis of the finite-size scaling   relations (\ref{eq:correlationratio}) and (\ref{eq:susceptibility}) without taking  the correction to scaling terms ($L^{-\omega}$) into consideration. As expected and as shown in Figs.~\ref{fig:datacollapse}(a) and (b), the data for different system sizes  collapse onto each other in the  \textit{large}   size limit.

\subsection{Single particle Green's functions}

To extract the
fermionic anomalous dimension, we   consider the imaginary time displaced local  single particle  Green's function  at $\tau =  \beta/2 \equiv L/2$:
\begin{equation}\label{Eq:Green}
  G(\lambda,L)  =\frac{1}{2L^2}\sum_{{\ve r},\delta,\sigma}
   \langle \hat{{c}}^{\dagger}_{{\ve r}+\ve{\delta}, \sigma}(\beta/2)\hat{{c}}_{{\ve r}+ \ve{\delta}, \sigma}(0) \rangle.
\end{equation}
Here  $\ve{r}$ denotes the unit cell, $\ve{\delta}$ is the orbital in the unit cell corresponding to the A(B) sublattices, and $\sigma$ is the spin  degree of freedom.
It is    convenient  to normalize $G(\lambda,L)$  with its non-interacting value so as to filter out size effects.
This motivates the   definition:
\begin{equation}\label{Eq:Zr}
   Z =  \frac{G(\lambda,L)}{G(0,L)}.
\end{equation}
In the non-interacting case, $G(0,L)$  scales as $L^{-2}$   reflecting the   fermionic  anomalous dimension, $d/2$  ($d$ is the  spatial dimension),   of the
fermion operator  at the non-interacting fixed point  (see Appendix~\ref{AppendixB}  for a symmetry based  discussion of the single particle Green's function).

In the vicinity of the  GN critical point, we expect:
\begin{equation}\label{eq:particleweight}
Z= L^{-\eta_{\psi}}f_{Z}\left( L^z/\beta,\left( \lambda - \lambda_c \right)L^{1/\nu}, L^{-w}  \right),
\end{equation}
where $\eta_{\psi}$  is the
fermionic  anomalous dimension.    In Fig.~\ref{fig:SPGreen}(a)  we  report our bare data from which  we can extract
$\eta_{\psi}$   using  the relation:
\begin{equation}
\label{eq:eta_psi}
\eta_{\psi}(L,rL)=-\frac{1}{\ln(r)}\ln \left
( \frac{Z(\lambda_c(L),rL)}{Z(\lambda_c(L),L)}\right)
\end{equation}
with $r = \frac{L}{L+6}$,  and $\lambda_c(L)$ the   size resolved crossing points of the correlation ratio.
In Fig.~\ref{fig:SPGreen}(b)  we show that  $\eta_{\psi}=0.29(2)$ with  $\eta_{\psi}(L)=\eta_{\psi}+d_1L^{-\omega_4}$  and $\omega_4=3.2(6)$.
We note that the single particle Green's function is not a Lorentz invariant quantity (see Appendix\ref{AppendixB}).  It is hence challenging to  use the  real space decay  so as to  extract the fermion anomalous dimension.

\section{Discussions and outlook}
\label{sec:conclusions}
For $N_f=2$ four component Dirac  fermions  akin to graphene, there are a number of GN transitions    that can be classified in terms of symmetry.
After a canonical  transformation,  the non-interacting Dirac Hamiltonian of graphene  is given by (see Appendix \ref{AppendixB}),
\begin{equation}
\label{Dirac.eq}
  \hat{H}_0  =   -v_F  \sum_{\ve{p}}  \hat{\ve{\Psi}}^{\dagger}(\ve{p})      \left[  p_x \tau_{x}      + p_y \tau_{y} \right]
 \hat{\ve{\Psi}}^{\phantom\dagger}(\ve{p}),
\end{equation}
where  we  label  the eight-component spinor as $ \hat{\ve{\Psi}}^{\dagger}  :=  \Psi^{\dagger}_{\tau,\sigma,\mu} $. The  $\tau_{x,y,z}$   Pauli  matrices act on the  $\tau$ indices and a similar  notation holds for $\sigma_{x,y,z}$   and $\mu_{x,y,z}$  Pauli  matrices.     In this writing of the  Dirac  Hamiltonian, the SU(4)   symmetry  is  explicit.
$\hat{H}_0 $  has a   maximum of  five   mutually anti-commuting mass terms  corresponding to the matrices:
\begin{equation}
	\ve{\Gamma}    = \left(   \ve{\sigma} \mu_x  \tau_z,  \mu_y\tau_z,  \mu_z \tau_z \right).
\end{equation}
The GN models
\begin{equation}	
	 \hat{H}_N =   \hat{H}_0   +     U  \sum_{i=1}^{N}        \int_{V} d^2 \ve{x}   \left(   \hat{\Psi}^{\dagger}(\ve{x})  \Gamma_i  \hat{\Psi}^{\phantom\dagger}(\ve{x})  \right)^2
\end{equation}
have  O(N)  symmetry,  and the generators of the SO(N) sub-group are  given by:
\begin{equation}
     \Gamma_{ij} =  \frac{i}{4} \left[   \Gamma_i,  \Gamma_j\right]  \;   \;        i  > j
\end{equation}
where   $ i \in 1 \cdots N $.    The authors of Ref.~\cite{Janssen18}   compute   within  an $\epsilon$ expansion around three spatial dimensions, as well as  with functional renormalization group (FRG)  methods the exponents for the aforementioned O(N)-GN transitions.     In the FRG approximation,  the bosonic anomalous dimensions  read:   $\eta_{\phi} =   0.760, 0.875, 1.015, 1.159,  $ and $ \eta_{\phi} = 1.285$    at  $N=1,2,3,4,5$  respectively.    Hence,  as  $N$ grows
it becomes  increasingly important to compute  susceptibilities   rather than equal time correlation functions.     Lattice  regularizations of  the above continuum theories  can capture the   O(1) or  $Z_2$  \cite{He17},   O(2)   \cite{Li17,Otsuka18}  as well as the O(3)    \cite{Assaad13,Toldin14,Otsuka16,Otsuka20}  critical points.
While Landau level regularization schemes allow to simulate higher symmetries \cite{Ippoliti18,WangZ20}, O(4) and O(5) Gross-Neveu transitions seem to be realized only at multi critical points \cite{Janssen18,Roy19,Torres19}. Such multi  critical points have been put forward in  fermion lattice models  in  Refs.~\cite{SatoT17,sato2020topological}  and Ref.~\cite{li2019deconfined}  for the O(4) and O(5) cases respectively.   Aside for the necessity of considering susceptibilities to investigate criticality the task  becomes especially challenging   since one has to control two model parameters to locate the critical point.

\begin{table}[h]
\vspace*{0.5cm}
\begin{center}
 \begin{tabular}{|| l|c  |c   | c ||}
 \hline
      &  $1/\nu$ & $\eta_{\phi}$   & $\eta_{\psi}$     \\ [0.5ex]
 \hline\hline
 This study &  1.11(4) & 0.80(9)   &  0.29(2) \\
 \hline Ref.~\cite{Otsuka20}   (AFQMC) & 0.95(5)  & 0.75(4)   &  0.23(4)\\
 \hline Ref.~\cite{Buividovich18}  (HMC)  & 0.861 & 0.872(22) &  ---  \\
 \hline Ref.~\cite{Liu18} (AFQMC) & 1.14(9)  & 0.79(5) & ---- \\
 \hline Ref.~\cite{Otsuka16} (AFQMC)  & 0.98(1) & 0.49(2) & 0.20(2) \\
  \hline Ref.~\cite{Toldin14} (AFQMC) & 1.19(6)  & 0.70(15) & ---- \\
\hline Ref.~\cite{Zerf17}  $(4-\epsilon) $,  $\epsilon^4$,  Pad\'e [2/2]&  0.6426 &  0.9985 &  0.1833 \\
\hline Ref.~\cite{Zerf17}  $(4-\epsilon) $,  $\epsilon^4$,  Pad\'e [3/1]&  0.6447 &  0.9563 &  0.1560 \\
\hline Ref.~\cite{Knor18}  FRG &   0.795 &  1.032&  0.071 \\
\hline Ref.~\cite{Janssen14a} FRG &  0.76&  1.01 & 0.08 \\
   [1ex]
 \hline
\end{tabular}
\end{center}
\caption{ \label{Table} Comparison of critical  exponents of the $N_f =2 $ four-component Dirac fermions Gross-Neveu  O(3)   critical point in 2+1 dimensions.
The table is adapted from Ref.~\cite{Huffman19}.    }
\end{table}
In  Hubbard based models, generically used to  capture  GN O(3)  criticality, computing the   susceptibilities of the   bosonic mode turns out to be difficult  to compute  due to anomalous fluctuations  that suggest fat tailed  distributions.
When computing observables in  the AFQMC, we  divide by the fermion determinant \cite{Assaad08_rev}.  The zeros of this quantity could  be at  the origin of these anomalous fluctuations.  This  interpretation has been put forward in Ref.~\cite{Hao16}.   It certainly may be part of the problem,  but does not  seem to provide an understanding of why the spin-susceptibility shows anomalous fluctuations but not,  for instance, the charge susceptibility or  the  single particle time displaced  correlation  function.  We refer the reader to Appendix~\ref{AppendixA}  for  further discussions and examples.

We have noticed empirically that   the AFQMC implementation of the model of Eq.~\ref{Eq:Ham}  \cite{Liu18} showing a  GN O(3)  transition from  a DSM to a QSH  insulator does not suffer  from the aforementioned issue.   It hence provides a unique  possibility to  compute the exponents by considering susceptibilities rather than equal time correlations.     Our  results are at best summarized by comparing with  other calculations listed in Table \ref{Table}. The Monte Carlo results are  ordered  chronologically  and convergence between different groups is apparent.  In particular, the most recent  independent    calculations  of  Ref.~\cite{Otsuka20},     where   the  Dirac metal  originates form a d-wave superconducting BCS state  and the antiferromagnetic  mass terms  are generated dynamically   with a Hubbard U term, compare very  favorably to our DSM to QSH transition.

To progress in our determination of the critical exponents,  high precision simulations on larger system sizes are  desirable.    In AFQMC algorithms, the  fermion determinant is computed exactly  such that  the  computational time per sweep   for  $\beta = L$ scaling   reads $L^7$.      Alternatively,  in hybrid Monte Carlo  (HMC)   approaches \cite{Duane87,Beyl17}  one generically  evaluates the  fermion determinant stochastically   such that one can,   in the ideal case,  hope for an $L^4$  scaling corresponding to the  Euclidean volume.    In the vicinity of the GN critical point, such a scaling is not achievable,  and the authors of Ref.~\cite{Buividovich18}   revert to an  explicit calculation of the fermion determinant \cite{Ulybyshev19b}. The origin for this poor scaling of the HMC,   are  zeros of the fermion determinant.   As mentioned above, one can conjecture that our ability to compute the  order parameter susceptibilities stems from a low density of zeros of the fermion determinant. If so, it may be  worth while to attempt HMC simulations of our model   in the hope of reaching larger system sizes.

\begin{acknowledgments}
We would like to thank Y. Otsuka, K. Seki, S. Sorella, F. Parisen Toldin,  M. Ulybyshev ,S. Yunoki and Disha Hou for
valuable   discussions.     The
authors gratefully acknowledge the Gauss Centre for Supercomputing
e.V. (www.gauss-centre.eu) for funding this project by providing computing
time on the GCS Supercomputer SUPERMUC-NG at Leibniz Supercomputing Centre
(www.lrz.de).  FFA thanks the W\"urzburg-Dresden Cluster of Excellence on
Complexity and Topology in Quantum Matter ct.qmat (EXC 2147, project-id
390858490). Z.W. thanks financial support from the DFG funded SFB 1170
on Topological and Correlated Electronics at Surfaces and Interfaces.
T.S. thanks funding from the Deutsche Forschungsgemeinschaft under the grant number SA 3986/1-1.
Y.L. was supported by the
China Postdoctoral Science Foundation under Grants
No.2019M660432 and No.2020T130046 as well as the National Natural Science
Foundation of China under Grants No.11947232 and
No.U1930402.
W.G. was supported by the National Natural Science Foundation
of China under Grants No. 11775021 and No. 11734002.
\end{acknowledgments}

\appendix
\section{ Time displaced correlation functions }
\label{AppendixA}
In this   appendix we present simulations for the  Hubbard model on  the honeycomb lattice close to the GN O(3)  critical point.  Our aim is to illustrate the difficulty in computing precisely the  time displaced spin-spin correlations.   In contrast  the   corresponding data for the model of Eq.~\ref{Eq:Ham}    shows no such anomalous   fluctuations up to  $L=\beta=24$.

\begin{figure}[h]
   \includegraphics[width=0.45\textwidth]{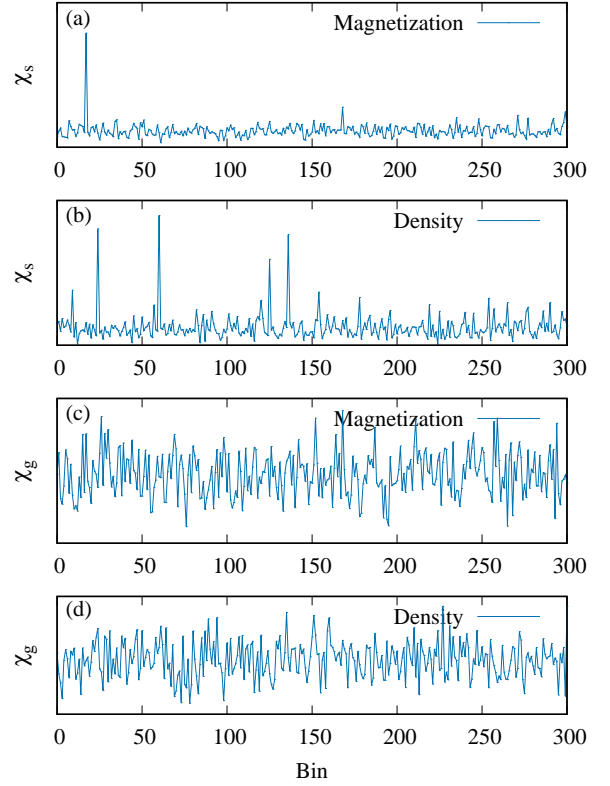}
   \caption{
Bin values for the  spin (a), (b) and single particle (c), (d) susceptibilities for a $12\times 12$ honeycomb lattice at
   $U/t=4$ and  $\beta t = 12$. (a) , (c) The HS  field couples  to the magnetization. (b), (d)  The HS field couples to the density.  Each bin  consists of  2400 sweeps. }
    \label{fig:HisttU}
\end{figure}

Using  the ALF-2.0 library \cite{alfcollaboration2021alf,Assaad08_rev},   we can choose between different  Hubbard Stratonovich  (HS) transformations:  the  field can couple to the density or to the magnetization \cite{Hirsch83}. The density decoupling is an SU(2)  spin invariant code, meaning that for each field configuration,   global SU(2)  spin symmetry is  present.
\begin{figure}[h!]
    \includegraphics[width=0.45\textwidth]{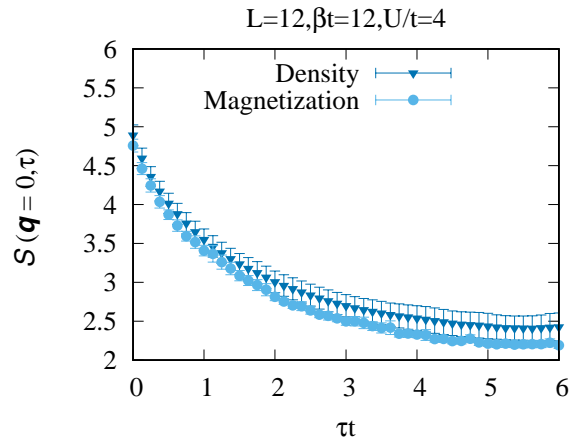}
                \caption{ Spin-spin time displaced correlation  function  at the ordering wave vector. Here we  consider    the   Hubbard model on the honeycomb lattice in the proximity of the GN O(3) critical point.    We  present data for  different choices of the HS transformation    where the field couples to the density (triangles) or to the magnetization (circles). }
        \label{tU_spin.fig}
\end{figure}
On the other  hand,  coupling to the  magnetization breaks the SU(2) symmetry  to U(1).  This symmetry will be restored  after sampling over   auxiliary field configurations.   In  Fig.~\ref{tU_spin.fig}    we plot the spin-spin correlations,
\begin{equation}
	S(\ve{q},\tau)    =    \frac{4} {3}\sum_{\delta}    \sum_{\ve{r}}  e^{i \ve{q} \cdot \ve{r} } \langle  \ve{S}_{\ve{r},\delta}(\tau) \ve{S}_{\ve{0},\delta}(0)  \rangle,
\end{equation}
where   $\ve{r}$  denotes a unit cell, $\delta$ the orbital  and  $\ve{S}_{\ve{r},\delta}$  is the spin operator.   Here we consider an $L=12$ lattice at $\beta t = 12$.   As apparent, and within error-bars, both  HS transformation yield identical results.
\begin{figure}[h!]
   \includegraphics[width=0.45\textwidth]{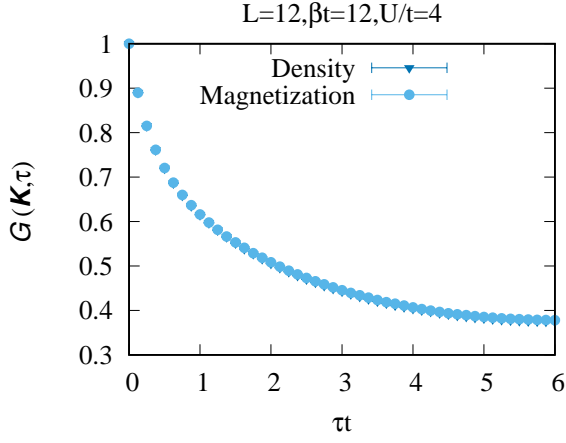}
   \caption{Green's function at the Dirac point for the same run as in Fig.~\ref{tU_spin.fig} }
    \label{tU_green.fig}
\end{figure}
To assess the   quality of the data we  plot in Fig.~\ref{tU_green.fig} the single particle Green's function:
\begin{equation}
   G(\ve{k},\tau)   = \frac{1}{2} \sum_{\delta,\sigma}   \langle  \hat{c}^{\phantom\dagger}_{\ve{k},\delta,\sigma} (\tau)   \hat{c}^{\dagger}_{\ve{k},\delta,\sigma}  \rangle
\end{equation}
for the same run that produced the data of Fig.~\ref{tU_spin.fig}.   As apparent the quality of the single particle Green's function is excellent  in comparison to the  time displaced spin correlations. The larger error bars observed in  the  spin  channel stem from  \textit{ rare }  configurations with  \textit{  anomalous } fluctuations.     The values of each bins  for the spin
\begin{equation}
	 \chi_s  = \int_{0}^{\beta} d \tau S(\ve{q}=0,\tau)
\end{equation}
and  single particle
\begin{equation}
	 \chi_g  = \int_{0}^{\beta} d \tau G(\ve{k}=0,\tau)
\end{equation}
susceptibilities are  plotted in Fig.~\ref{fig:HisttU}.
For the spin susceptibilities,  one observes  \textit{ spikes }   in the bin values  for both codes. On the other hand  the bin values of the Green's function susceptibility shows no  anomalies.

\begin{figure}[h!]
\begin{center}

  \includegraphics[width=0.45\textwidth]{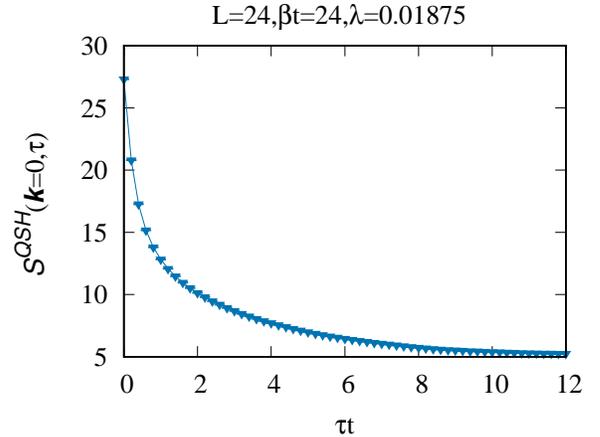}
  \end{center}
  \caption{ Time displaced spin-orbit correlation functions  at  $L = \beta= 24 $ and $\lambda = 0.01875$. \label{fig:QSH_K0} }
\end{figure}

\begin{figure}[h!]
\begin{center}
  \includegraphics[width=0.45\textwidth]{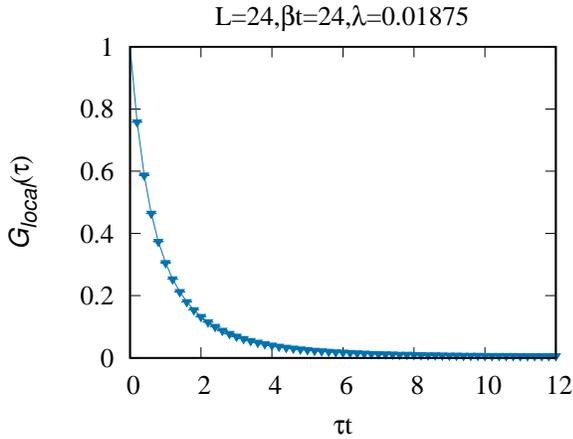}
  \end{center}
  \caption{ Time displaced local Green's function for the same run
as in Fig.~\ref{fig:QSH_K0} \label{fig:Green_R0} }
\end{figure}

 \begin{figure}[h]
   \includegraphics[width=0.45\textwidth]{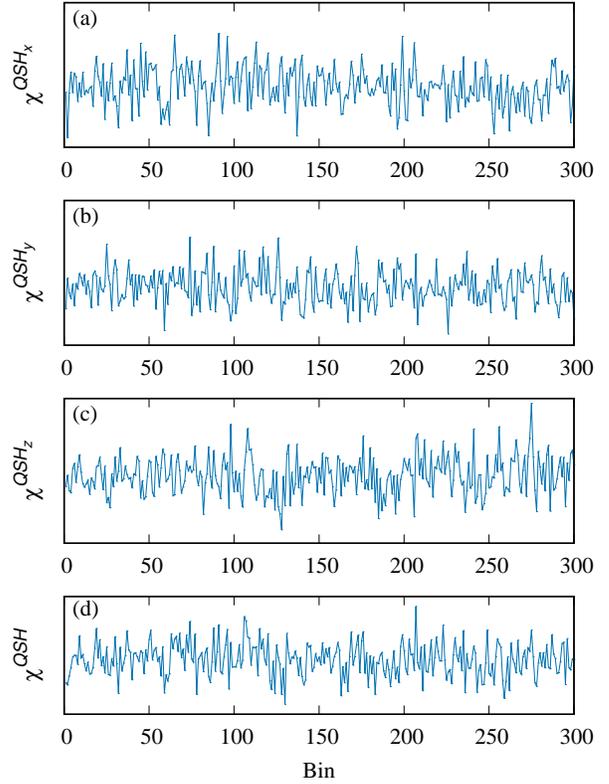}
   \caption{   Bin values for spin-orbital coupling susceptibility $\chi^{QSH}$ (d) and its three components $\chi^{QSH_x}$ (a), $\chi^{QSH_y}$(b) and $\chi^{QSH_z}$(c) for a $12\times 12$ honeycomb lattice at
   $\lambda/t=0.0186$ and  $\beta t = 12$. Each bin  consists of  2400 sweeps.}
    \label{fig:QSH_Bin}
\end{figure}

We now consider equivalent quantities albeit on much larger system sizes for the model of Eq.~\ref{Eq:Ham}.  In  Fig.~\ref{fig:QSH_K0}  we plot,
\begin{eqnarray}
&&S^{QSH}({\ve k}, \tau)=\nonumber\\
&& \sum_{ \langle\langle\ve{\delta},\ve{\delta}'\rangle\rangle}\sum_{ {{\ve r}} } {\ve e}^{i {\ve k}\cdot{{\ve r}}}   \left<  \hat{\ve{O}}^{QSH}_{ {\ve r},\langle\langle\ve{\delta},\ve{\delta}'\rangle\rangle}(\tau)\cdot \hat{\ve{O}}^{QSH}_ { {\ve 0},\langle\langle\ve{\delta},\ve{\delta}'\rangle\rangle}(0) \right> \,
\end{eqnarray}
where $ \hat{\ve{O}}^{QSH}_ { {\ve r},\langle\langle\ve{\delta},\ve{\delta}'\rangle\rangle}$ is defined in Eq.~\ref{eq:O_QSH}.
As  apparent, the data is of excellent  quality.

In Fig.~\ref{fig:Green_R0}    we plot the local Green's function
\begin{equation}\label{Eq:Greenr0}
  G_{\text{local}}(\tau)  =\frac{1}{2L^2}\sum_{{\ve r},\delta,\sigma}
   \langle \hat{{c}}^{\dagger}_{{\ve r}+\ve{\delta}, \sigma}(\tau)\hat{{c}}_{{\ve r}+ \ve{\delta}, \sigma}(0) \rangle.
\end{equation}
used to obtain the  fermion anomalous dimension.  As  apparent the data  quality is very good.

The spin-orbital coupling susceptibility reads,

\begin{equation}
   \chi^{QSH} = \Lambda_1 \left( \mbox{$\int_{0}^{\beta}$}  d\tau S^{QSH}_{\langle\langle\ve{\delta},\ve{\delta}\rangle\rangle, \langle\langle\ve{\delta}'',\ve{\delta}'''\rangle\rangle}({\ve k=0},\tau)  \right) \,,
\end{equation}
where,$S^{QSH}_{\langle\langle\ve{\delta},\ve{\delta}\rangle\rangle, \langle\langle\ve{\delta}'',\ve{\delta}'''\rangle\rangle}$ is the time displaced correlation function of spin orbit coupling order parameter and  $\Lambda_1()$ indicates the largest eigenvalue of the  $6 \times 6$ matrix spanned by the next-nearest neighbor bonds of a hexagon.
The values of each bins for $\chi^{QSH}$ and  the three components $\chi^{QSH_x}$ , $\chi^{QSH_y}$ and $\chi^{QSH_z}$ are plotted in Fig.\ref{fig:QSH_Bin}.  We observe no  \textit{ spikes } in the bin values  for all  components.

\section{  Space and time  dependence of the single particle Green's function }
\label{AppendixB}

The aim of this appendix  is to  understand  the  behavior of the single particle Green's function  in space and imaginary time   using symmetry  arguments. Let us start with the tight binding Hamiltonian on the honeycomb  lattice that reads,
\begin{equation}	
   \hat{H}_0 = -t \sum_{\ve{k} \in BZ}    \left( \hat{a}^{\dagger}_{\ve{k}}, \hat{b}^{\dagger}_{\ve{k}} \right)
   \begin{pmatrix}
0     & Z(\ve{k})  \\
\overline{Z(\ve{k})} & 0
\end{pmatrix}
\begin{pmatrix}
\hat{a}^{\phantom\dagger}_{\ve{k}} \\
\hat{b}^{\phantom\dagger}_{\ve{k}}
\end{pmatrix}.
\end{equation}
Here,
\begin{equation}
	\hat{a}^{\dagger}_{\ve{k}} = \frac{1}{\sqrt{N}}   \sum_{\ve{r}} e^{i \ve{k} \cdot \ve{r}} \hat{a}^{\dagger}_{\ve{r}}
\end{equation}
creates a Bloch  state   on  orbital $a$ of the unit cell.    A similar equation holds for  the $b$-orbital.  $\ve{r}   = n \ve{a}_1   + m \ve{a}_2 $   with  $ \ve{a}_1 = a (1,0) $,
$\ve{a}_2  = a \left( \frac{1}{2}, \frac{\sqrt{3}}{2} \right) $   and $n,m \in \mathbb{Z}$  runs over the unit cells.    We  have used periodic boundary conditions and
\begin{equation}
	 Z(\ve{k})  =  1 + e^{- \ve{k}\cdot \ve{a}_2}   +   e^{- i \ve{k} \cdot \left(  \ve{a}_2 - \ve{a}_1 \right) }.
\end{equation}
The Dirac  points are defined by the zeros of $Z(\ve{k})$ and are located at:
\begin{equation}
	  \ve{k} =  \pm \ve{K},  \text{ with }  \ve{K} =  \frac{4}{3} \ve{b}_1  + \frac{2}{3} \ve{b}_2
\end{equation}
with  $\ve{b}_i \cdot \ve{a}_j = 2 \pi \delta_{i,j} $.

The   Hamiltonian is  invariant under the   anti-unitary particle-hole  transformation
\begin{equation}
\hat{T}^{-1}
 \alpha
\begin{pmatrix}
\hat{a}^{\dagger}_{\ve{r}}  \\
\hat{b}^{\dagger}_{\ve{r}}
\end{pmatrix}
\hat{T}   =
\overline{\alpha}
\begin{pmatrix}
\hat{b}^{\phantom\dagger}_{\ve{r}}  \\
-\hat{a}^{\phantom\dagger}_{\ve{r}}
\end{pmatrix}
\end{equation}
as  well as  under inversion symmetry,
\begin{equation}
\hat{I}^{-1}
\begin{pmatrix}
\hat{a}^{\dagger}_{\ve{r}}  \\
\hat{b}^{\dagger}_{\ve{r}}
\end{pmatrix}
\hat{I}  =
\begin{pmatrix}
\hat{b}^{\dagger}_{-\ve{r}}  \\
\hat{a}^{\dagger}_{-\ve{r}}
\end{pmatrix}.
\end{equation}
Hence, for $\ve{r} \ne \ve{0} $,
\begin{eqnarray}
& & \langle \hat{a}^{\dagger}_{\ve{r}}  \hat{a}^{\phantom\dagger}_{\ve{0}} \rangle    = \langle \hat{b}^{\phantom\dagger}_{\ve{r}}  \hat{b}^{\dagger}_{\ve{0}} \rangle   =  - \langle \hat{b}^{\dagger}_{\ve{0}}  \hat{b}^{\phantom\dagger}_{\ve{r}}  \rangle  = \nonumber \\
- & &  \langle \hat{b}^{\dagger}_{-\ve{r}}  \hat{b}^{\phantom\dagger}_{\ve{0}}  \rangle   =
 -\langle \hat{a}^{\dagger}_{\ve{r}}  \hat{a}^{\phantom\dagger}_{\ve{0}}  \rangle
\end{eqnarray}
and  $ \langle \hat{a}^{\dagger}_{\ve{r}}  \hat{a}^{\phantom\dagger}_{\ve{0}} \rangle $  vanishes.    In the last   two steps, we have used   translation and inversion symmetry.
Similarly,  one will show that, $  \langle \hat{b}^{\dagger}_{\ve{r}}  \hat{b}^{\phantom\dagger}_{\ve{0}} \rangle =0$ again for $\ve{r} \ne \ve{0} $.   Hence,  provided that  the symmetries of the  Dirac Hamiltonian are not broken,  only equal time correlations between  different orbitals   do not vanish.

We will now  show that there is no non-vanishing Lorentz invariant fermion bi-linear  such that we cannot expect a \textit{simple} asymptotic  behavior  of the one particle propagator.
\begin{figure}[b]
  \includegraphics[width=0.45\textwidth]{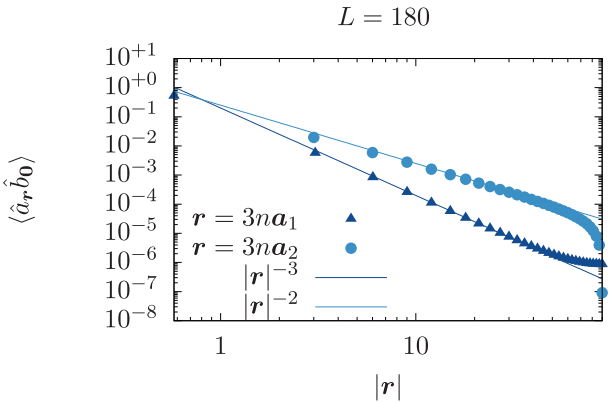}
  \caption{\label{fig:Greenr_U0} Real space   equal time  Green's function  at zero temperature  along different directions. }
\end{figure}
Since Lorentz symmetry is emergent, we will consider the  continuum  limit by expanding around  the Dirac points:
 \begin{eqnarray}
	& & Z(\phantom{-}\ve{K}  +  \ve{p } )    = \phantom{-} \frac{\sqrt{3}a}{2} \left(  p_x  - i   p_y \right),    \nonumber \\
	& & Z(- \ve{K}  +  \ve{p } )    =  - \frac{\sqrt{3}a}{2} \left(   p_x  + i  p_y \right),
\end{eqnarray}
to obtain:
\begin{equation}
    \hat{H}   = - v_F  \sum_{\ve{p}, i= 1,2} \hat{\ve{c}}^{\dagger}_{\ve{p}}  i p_i \gamma_0  \gamma_i \hat{\ve{c}}^{\phantom \dagger}_{\ve{p}}.
\end{equation}
Here  $ \hat{\ve{c}}^{\dagger}_{\ve{p}}  \equiv \hat{c}^{\dagger}_{\ve{p}, \mu= \pm \ve{K},   \tau = a,b} $      such that
$ \hat{c}^{\dagger}_{\ve{p}, \mu =  \pm \ve{K},   \tau = a } =   \hat{a}^{\dagger}_{\pm \ve{K} + \ve{p} }  $  and
$ \hat{c}^{\dagger}_{\ve{p}, \mu =  \pm \ve{K},   \tau = b } =   \hat{b}^{\dagger}_{\pm \ve{K} + \ve{p} }  $.   The Fermi velocity is given by  $v_F  = \frac{\sqrt{3}a t }{2} $ and the  $\gamma$-matrices are defined as
\begin{equation}
\gamma_0 =   \tau_z, \gamma_1 =   \mu_z  \tau_y ,  \gamma_2 =\tau_x,   \gamma_3= \mu_x \tau_y ,  \gamma_5  =  \mu_y  \tau_y.
\end{equation}
 $\ve{\tau} $  and $ \ve{\mu} $  are vectors of Pauli spin matrices that act on orbital and valley indices respectively.     As  apparent  the $\gamma$-matrices   satisfy the Clifford algebra,
 \begin{equation}
     \left\{ \gamma_{\mu},  \gamma_{\nu}  \right\}  = 2 \delta_{\mu,\nu}.
 \end{equation}
 Note that the canonical transformation that leads to Eq.~\ref{Dirac.eq}  is given by:
 \begin{equation}
	\hat{\ve{c}}^{\dagger}  =  \hat{\Psi}^{\dagger} \left(  \tau_{y} P_{+}  +  P_{-} \right)  \text{  with }   P_{\pm} = \frac{1}{2} \left( 1 \pm \mu_z \right).
\end{equation}
With
\begin{equation}
     \hat{\ve{c}}^{\dagger}_{\ve{p}}  =  \frac{1}{\sqrt{V}} \int_V  d^2{\ve{x}}  e^{ i \ve{p} \cdot \ve{x} }     \hat{\ve{c}}^{\dagger}(\ve{x}),
\end{equation}
the Euclidean time  action  is then given by:
\begin{equation}
S  =   v_F  \int   d^{2}\ve{x}  d \tau    \sum_{\mu=0,2}  \overline{\ve{c}}^{\phantom\dagger}(\ve{x})   \partial_{\mu}   \gamma_{\mu}
\ve{c}^{\phantom\dagger}(\ve{x}).
\end{equation}
In the above,
\begin{equation}
 \overline{\ve{c}}^{\phantom\dagger}(\ve{x})  = \ve{c}^{\dagger}(\ve{x})  \gamma_0,
\end{equation}
\begin{equation}
\nonumber
   \partial_0 =  \frac{\partial}{v_F \partial \tau }   \text{ and } \, \,   \partial_i =  \frac{\partial}{ \partial x_i}
\end{equation}
and $\ve{c}_{\ve{p}} $ is a  Grassmann spinor.
The Dirac equation is scale invariant.  In particular  under the transformation $  \ve{x}' =  b \ve{x} $  and $ \tau'  = b \tau $   the Euclidean action remains  form  invariant   provided  that  the  fermion fields transform as
\begin{equation}
\ve{c}'^{\phantom\dagger} (\ve{x}')    = b^{-d/2}  \ve{c}^{\phantom\dagger}(\ve{x})
\end{equation}
for the  two, $d=2$,  dimensional case.     Hence,   fermion bilinears   can take the form:
\begin{equation}
    \langle  \overline{\ve{c}}(x)   M  \ve{c}(0)  \rangle    \propto  \frac{a}{(v_F \tau)^2}   +   \frac{b}{|\ve{x}|^2  }  +  \frac{c x_1}{|\ve{x}|^3} +
 \frac{d x_2}{|\ve{x}|^3} + \frac{a_L}{(v_F \tau)^2 +  |\ve{x}|^2}\cdots
\end{equation}
The Dirac equation  is Lorentz invariant  \cite{Peskin_book}  such that Lorentz invariant  fermion bi-linears   scale as
\begin{equation}
    \langle  \overline{\ve{c}}(x)   M_{L}  \ve{c}(0)  \rangle    \propto  \frac{a_L}{(v_F \tau)^2 +  |\ve{x}|^2}.
\end{equation}
Example of Lorentz invariant bilinears include
\begin{equation}
	M_L  = 1, \,  M_L  = i\gamma_3,  \,   M_L = i \gamma_5,  \,   M_L  =     i \gamma_3 \gamma_5.
\end{equation}
These  biliniears     are mass terms corresponding  respectively to  charge-density wave (CDW)  patterns,  to the  two K\'ekule orders and  finally  to the  Haldane mass.   Since  mass terms break symmetries of the Dirac  Hamiltonian they vanish such that
\begin{equation}
\langle  \overline{\ve{c}}(x)   M_{L}  \ve{c}(0)  \rangle   =0.
\end{equation}
One can check  the above explicitly for the CDW  mass since it   changes sign under inversion symmetry.
We are  hence left with fermion bi-linears that are not Lorentz invariant,  and hence do not enjoy rotational symmetry in space and imaginary time.  In particular computing $ \langle \hat{a}^{\dagger}_{\ve{r}} \hat{b}_{\ve{0}} \rangle $  on the lattice amounts to considering
$ M = \gamma_0   \gamma_2 $.  This is a nematic term that breaks Lorentz symmetry.
An explicit   calculation  of the equal time  correlations  of this fermion bilinear can be found in an appendix of Ref.~\cite{Seki19}.   For  distances on the lattice   that satisfy   $\ve{r}   = n 3 \ve{a}_1  + m 3 \ve{a}_2 $,  $ e^{ \pm i \ve{K} \cdot \ve{r} } = 1 $ and no oscillatory behavior is seen.  In Fig.~\ref{fig:Greenr_U0}    we plot the equal  time Green's function  using these sets of points.  As apparent,  depending upon the direction $1/r^2$  and  $1/r^3$    decays  are observed.    Note that the $1/r^3$  decay   can be justified   by   combining    the terms  $ x_1/ |\ve{x}|^3 $ and $ x_2/ |\ve{x}|^3 $   for  $ x_1  =  x $ and $ x_2    =  a - x$.

Setting  $\ve{x} =0 $  and  considering solely  imaginary time,  greatly  simplifies the analysis.  In this case the scaling dimension of the fermion leads to
\begin{equation}
	 \langle \hat{a}^{\dagger}_{\ve{0}} (\tau)  \hat{a}^{\phantom\dagger}_{\ve{0}} (\tau=0) \rangle   \propto \frac{1}{(v_F \tau)^2}.
\end{equation}
In fact an explicit calculation of this quantity  on the lattice  and at  zero temperature   gives:
\begin{equation}
\langle \hat{a}^{\dagger}_{\ve{0}} (\tau)  \hat{a}^{\phantom\dagger}_{\ve{0}} (\tau=0) \rangle    =
\frac{1}{2N}  \sum_{\ve{k}} e^{- \tau   | t  Z(\ve{k})|}
\end{equation}
Expanding around the Dirac points,  $   Z(\pm \ve{K}  +  \ve{p} )   =  \frac{\sqrt{3}a}{2} |\ve{p}|  $  and changing sums to integrals,    yields
the   desired result.
\begin{figure}[b]
  \includegraphics[width=0.45\textwidth]{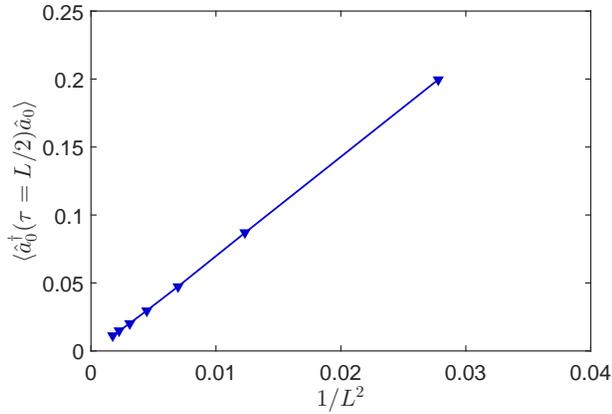}
 \caption{\label{fig:speqtau}
 Here we plot the time displaced local  Green's function, at $\tau =  \beta /2$  for $\beta =L $ simulations.   As apparent this  quantity is proportional to $L^{-2}$ .
 }
\end{figure}
Fig.~\ref{fig:speqtau}   shows that adopting a $\beta = L$ scaling    and considering  $\tau = \beta/2$   provides confirmation of the above law.

At the Gross-Neveu  critical point,     the scaling  dimension of the fermion operator  will be enhanced by half the fermion anomalous dimension, $\eta_{\Psi}$,  such that at this critical point we expect:
\begin{equation}
      \langle \hat{a}^{\dagger}_{\ve{0}} (\tau)  \hat{a}^{\phantom\dagger}_{\ve{0}} (\tau=0) \rangle_{GN}  \propto \frac{1}{(v_F \tau)^{2 + \eta_{\Psi} }}
\end{equation}
in two spatial dimensions.

\clearpage
\bibliography{./fassaad}
\end{document}